\documentclass[aps,epsfig,showpacs,twocolumn]{revtex4}

\usepackage{graphicx}

\begin{document}
\title{Adiabatic-Impulse approximation for avoided level crossings: 
from phase transition dynamics to  Landau-Zener evolutions  and back again
}
\author{Bogdan Damski and Wojciech H. Zurek}
\affiliation{
Theory Division, Los Alamos National Laboratory, MS-B213, Los Alamos, NM 87545, USA
}
\begin{abstract}
We show that a simple approximation based on concepts  underlying the
Kibble-Zurek theory of second order phase transition dynamics can be used to
treat avoided level crossing problems. The approach discussed in this paper
provides an intuitive insight into  quantum dynamics of  two level
systems, and may serve as a link between the theory of dynamics of 
classical and quantum  phase transitions. To illustrate these ideas we 
analyze dynamics of a paramagnet-ferromagnet quantum phase transition in the Ising model.
We also present exact unpublished  solutions of the Landau-Zener like problems.  
\end{abstract}
\pacs{03.65.-w,32.80.Bx,05.70.Fh}
\maketitle

\section{Introduction}

Two level quantum systems that undergo avoided level crossing 
play an important role in physics. Often they provide  not 
only a qualitative description of system properties but also
a quantitative one. The possibility for a successful two level approximation arises
in different physical systems, e.g., 
a single one-half spin; an atom placed in a resonant light;
smallest quantum  magnets; etc. 

In this paper we focus on avoided level crossing dynamics in  two level quantum
systems. Two interesting related extensions of above mentioned  long list 
are the critical dynamics of quantum phase transitions
\cite{subir,dorner,jacek,polkovnikov,ralf}
and adiabatic quantum computing \cite{ad_q_com}. In both of these cases (and,
indeed, in all of the other applications of the avoided level
crossing scenario) there are  many more than two levels, but the
essence of the problem can be still captured by the two level, Landau-Zener type,
calculation. 

Our interest in Landau-Zener model originates also from its well known numerous 
applications to different physical systems.
In many cases there is a possibility of  more general dependence
of the relevant parameters (i.e. gap between the two levels on time) than in 
the original Landau-Zener treatment. This motivates our extensions in this
paper of the level crossing dynamics to asymmetric level crossings and various power-law 
dependences. Appropriate   variations of external parameters 
driving Landau-Zener transition can allow for an experimental realization of
these generalized Landau-Zener like models in quantum magnets \cite{magnet}
and  optical lattices \cite{bloch}.

We focus  on  evolutions  that include both adiabatic and diabatic (impulse) regime 
during a single sweep of a system parameter. 
For simplicity, we call these evolutions diabatic 
since we  assume that they include a period of 
fast change of a system parameter. This is in  contrast to adiabatic time
evolutions induced by very  slow parameter changes when  the system never leaves 
adiabatic regime. 
We will  use a
formalism proposed recently by one of us \cite{bodzio}. It  
originates from the so-called Kibble-Zurek (KZ) theory of topological defect
production in the course of classical phase transitions \cite{kibble,zurek},
and works best for two level systems.
The two level approximation 
was recently  shown by Zurek {\it et al} \cite{dorner} and
Dziarmaga \cite{jacek} to be useful in studies of quantum phase transitions
confirming earlier expectations --  see Chap. 1.1 of \cite{subir}
and \cite{bodzio}.
Therefore,  we expect that the
formalism presented here will provide a link between dynamical 
studies of classical and quantum phase transitions. 

The present
contribution extends  the ideas presented in \cite{bodzio}. In particular, 
we have succeeded in replacing the fit to numerics used 
there  for getting a free parameter of the theory,
with a simple analytic calculation.
We also show that the method of
\cite{bodzio} can be successfully applied to a large  class of  two
level systems. Moreover,  we present nontrivial 
exact analytic solutions for Landau-Zener model and the relevant
adiabatic-impulse approximations not published to date.

Section \ref{aia} presents basics
of what we call the adiabatic-impulse approximation. Section \ref{LZ} shows
adiabatic-impulse, diabatic and finally exact 
solutions to  different versions of the Landau-Zener problem.
In Section \ref{2n} we discuss solutions for a whole class of two level systems
on the basis of adiabatic-impulse and diabatic schemes. 
Section \ref{ising} presents how results of Section \ref{LZ} can be used to study
quantum Ising model dynamics.
Details of
analytic calculations are  in Appendix \ref{a_l} (diabatic
solutions), Appendix \ref{a_lz} (exact solutions of Landau-Zener problems)
and Appendix \ref{a_r} (adiabatic-impulse solution from Sec. \ref{2n}).

\section{Adiabatic-Impulse approximation}
\label{aia}
We will study dynamics of quantum systems by 
assuming that it includes 
{\it adiabatic} (no population transfer between instantaneous energy eigenstates),
and {\it impulse} (no changes in the wave function except for an overall phase factor) stages
only.
The nature  of this approach suggests the name  {\it
adiabatic-impulse} (AI) approximation.
The AI approach
originates from the Kibble-Zurek (KZ) theory of nonequilibrium classical phase transitions
and we refer the reader to \cite{bodzio} for a detailed discussion of AI-KZ connections.
Below, we  summarize basic ideas of the KZ theory, and the
relevant assumptions used later on. 

Second order phase transitions and avoided level crossings share one key 
distinguishing characteristic: in both cases sufficiently near  the critical point 
(defined by the
relaxation time or by the inverse of the size of the gap between the two
levels)  system ``reflexes'' become very bad. 
In phase transitions this is known as ``critical slowing down''. 

In the treatment of the second order phase transitions this leads to the
behavior where the state of the system can initially -- in the adiabatic region
far away from the transition -- adjust to the change of the relevant parameter
that induces the transition, but sufficiently close to the critical point 
its reflexes become too slow for it to react at all. As a consequence, a
sequence of three regimes (adiabatic, impulse near the critical point, and
adiabatic again on the other side of the transition) can be anticipated.
In the second order phase transitions this allows one to employ a
configuration dominated by the  pre-transition fluctuations to calculate
salient features of the post-transition state of the order parameter 
(e.g., the density of topological defects). Key predictions of this KZ
mechanism  have been by now verified in numerical simulations \cite{antunes,laguna} and 
more importantly in the laboratory experiments \cite{kz1,kz2,kz3,kz4}.

The crux of this story  is of course the moments when the transition 
from adiabatic to impulse and back to adiabatic behavior takes place.
Assuming that the transition point is crossed at time $t=0$,
this must happen around the instants $\pm\hat t$, where 
 $\hat t$  equals system relaxation time $\tau$ \cite{zurek}.
As the relaxation time depends on system parameter $\varepsilon$, which
is changed as a function of time, one can write
\begin{equation}
\label{whz}
\tau(\varepsilon(\hat t\,))=\hat t.
\end{equation}
This basic equation proposed in \cite{zurek} can be solved when dependence of $\tau$ on some measure of
the distance from the critical point (e.g., relative temperature or coupling
$\varepsilon$) is known. Assuming, e.g., 
$\tau=\tau_0/\varepsilon$ and $\varepsilon= t/\tau_Q$,
where $\tau_Q$, the ``quench time", contains information about how fast the
system is driven through the transition, one arrives at
$\tau_0/(\hat{t}/\tau_Q)= \hat{t}$. Hence, 
the time $\hat{t}$ at which the behavior switches between approximately
adiabatic and approximately impulse is  $\sqrt{\tau_0\tau_Q}$.
Once this last ``adiabatic"  instant in the evolution of the system is
known, interesting features of the post transition state (such as the size of
the regions $\hat{\xi}$ in which the order parameter is smooth) can be
computed. Generalization to other universality classes is straightforward 
\cite{zurek} and leads to 
\begin{equation}
\hat{t}\sim \tau_Q^{1/(1+\nu z)} \ \ \ , \ \ \
\hat{\xi}\sim \tau_Q^{\nu/(1+\nu z)},
\end{equation}
where $z$ and $\nu$ are universal critical exponents (see, e.g., \cite{antunes}).

\begin{figure}
\includegraphics[width=\columnwidth, clip=true]{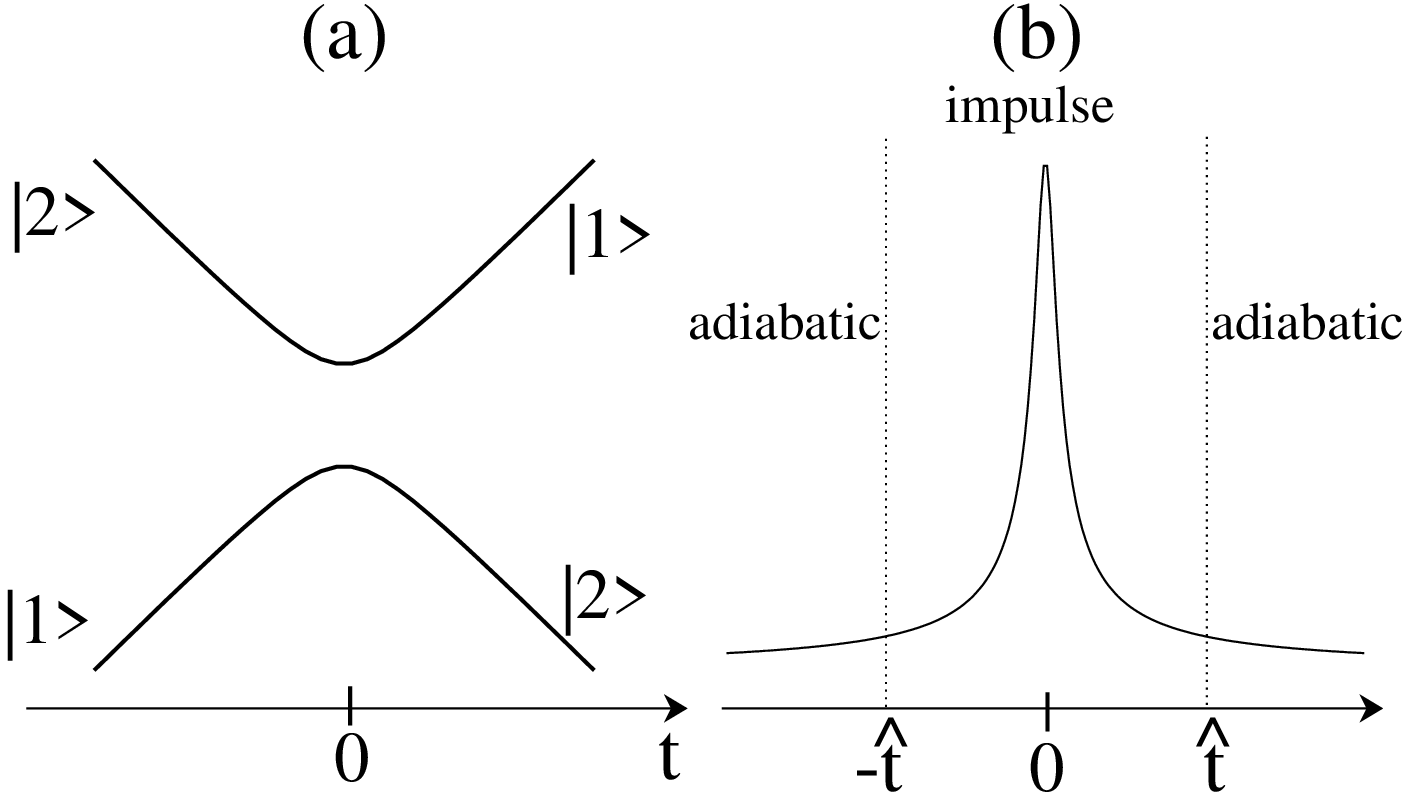}
\caption{Plot (a): structure of energy spectrum (parametrized by time) of a generic 
two-level system
under consideration. Note the anticrossing at  
$t=0$ and asymptotic form of eigenstates: $|1\rangle$, $|2\rangle$.
Plot (b): adiabatic-impulse regimes in  system dynamics. Compare it
to a plot of relaxation time scale in the Kibble-Zurek theory \cite{zurek,bodzio}. Note that 
instants separating adiabatic and impulse regimes does not have to be 
placed symmetrically with respect to $t=0$ -- see Sec. \ref{LZa} and Fig. \ref{asym}.
}
\label{gapanty}
\end{figure}

In what follows we adopt this approach to quantum systems where 
dynamics can be approximated as an avoided level crossing (see Fig.
\ref{gapanty}).
For simplicity  we
restrict ourselves to two-level systems that  possess a single 
anti-crossing (Fig. \ref{gapanty}) in the
excitation spectrum and a gap  preferably going to $+\infty$
far away from the anti-crossing. 
The latter condition guarantees that 
asymptotically the system enters adiabatic regime.

The passage through avoided level crossing 
can be divided into an adiabatic and an impulse regime according to the
size  of the gap in comparison with the energy scale that characterizes the rate of the 
imposed changes in system  Hamiltonian. If the gap is large enough 
the system is in adiabatic regime, while
when the gap is small it undergoes impulse time evolution as depicted in Fig.
\ref{gapanty}(b). The instants $\pm\hat{t}$  are supposed to be  such that 
the discrepancy  between time dependent exact results and
those coming from splitting of the evolution into only adiabatic and impulse parts
is  minimized. Therefore, the  AI approximation looks like  a
{\it time dependent variational method} where $\hat{t}$ is a variational parameter
while adiabatic-impulse assumptions provide a form of a variational wave
function.

To make sure that  assumptions behind the AI approach are 
correctly understood, let's consider time evolution of the system depicted
in Fig. \ref{gapanty}. Let the
evolution  start at $t_i\to-\infty$ from a ground state (GS) and 
last till $t_f\to+\infty$. The AI method assumes that the system wave
function, $|\Psi(t)\rangle$, satisfies the following three approximations  
coming from passing through first adiabatic, then impulse and finally 
adiabatic regions
\begin{eqnarray}
t\in[-\infty,-\hat{t}\,]&:&   
|\Psi(t)\rangle\approx({\rm phase \ factor})|{\rm GS \ at \ t}\rangle \nonumber\\
t\in[-\hat{t},\hat{t}\,]&:& |\Psi(t)\rangle\approx
({\rm phase \ factor})
|{\rm GS \  at  \ -\hat{t}}\rangle\nonumber\\
t\in[\hat{t},+\infty]&:&
|\langle\Psi(t)|{\rm GS  \ at \ t}\rangle|^2={\rm const}. \nonumber
\end{eqnarray}

Finally, one needs to know how to get the instant $\hat{t}$. As proposed in
\cite{bodzio}, the proper  generalization of Eq. (\ref{whz}) 
to the quantum case (after rescaling everything to dimensionless 
quantities) reads
\begin{equation}
\frac{1}{{\rm gap}(\hat{t}\,)}=\alpha\hat{t},
\label{bodzik}
\end{equation}
where $\alpha={\cal O}(1)$ is a constant. Similarity between Eq. (\ref{bodzik}) 
and Eq.  (\ref{whz}) suggests (in accord with physical intuition)  that the
quantum mechanical equivalent of the relaxation time scale is an inverse of the
gap.

In this paper we 
give a simple and  systematic way for (i) obtaining
$\alpha$ analytically; (ii) verification that Eq. (\ref{bodzik}) leads to  correct
results in the lowest nontrivial order. 
This method, illustrated on specific  examples in 
Appendix \ref{a_l}, is based on the observation that a time dependent
Schr\"odinger equation can be  solved exactly in the diabatic limit 
if one looks at the lowest nontrivial terms in  expressions for 
excitation amplitudes. This should be true  
even when getting an exact solution turns out to be very 
complicated or even impossible. After obtaining $\alpha$ this way, the whole AI
approximation is complete in a sense that there are no free parameters,
so its predictions can be rigorously checked  by comparison to exact (either 
analytic or numeric) solution.

\section{Dynamics in the Landau-Zener model}
\label{LZ}
In this section 
we  illustrate the AI approximation by considering the Landau-Zener model. In this way
we  supplement and extend the results of \cite{bodzio}. 

The Landau-Zener model, after rescaling all the quantities to dimensionless
variables, is defined by the Hamiltonian:
\begin{equation}
\label{H}
H=\frac{1}{2}
\left(
\begin{array}{cc}
\frac{t}{\tau_Q} & 1 \\
1 & -\frac{t}{\tau_Q}
\end{array}
\right),
\end{equation}
where $\tau_Q$ is time independent and provides a time scale on which the
system stays in the neighborhood of an anti-crossing. 
As $\tau_Q\to0$ the system undergoes diabatic time evolution, while
$\tau_Q\gg1$ means adiabatic evolution. 
There are two
eigenstates of this model for any fixed time $t$: 
the ground state $|\downarrow(t)\rangle$ and the excited state
$|\uparrow(t)\rangle$ given by:
\begin{equation}
\label{eig}
\left[
\begin{array}{c}
|\uparrow(t)\rangle \\ |\downarrow(t)\rangle 
\end{array}
\right] = 
\left(
\begin{array}{rr}
\cos(\theta(t)/2)  & \sin(\theta(t)/2) \\
-\sin(\theta(t)/2) & \cos(\theta(t)/2)
\end{array}
\right)
\left[
\begin{array}{c}
|1\rangle \\ |2\rangle 
\end{array}
\right],
\end{equation}
where $|1\rangle$ and $|2\rangle$ are time-independent basis states 
of the Hamiltonian (\ref{H}); $\cos(\theta)=\varepsilon/\sqrt{1+\varepsilon^2}$;
$\sin(\theta)=1/\sqrt{1+\varepsilon^2}$; 
\begin{equation}
\label{varepsilon}
\varepsilon= \frac{t}{\tau_Q},
\end{equation}
and $\theta\in[0,\pi]$. The gap in this model is
$$
{\rm gap}= \sqrt{1+\varepsilon^2}.
$$
The instant $\hat{t}$ is obtained from Equation (\ref{bodzik}):
$$
\frac{1}{\sqrt{1+\left(\frac{\hat{t}}{\tau_Q}\right)^2}}=\alpha\hat{t},
$$
which leads to the  following solution
\begin{equation}
\label{hatt}
\hat{\varepsilon}=\frac{\hat{t}}{\tau_Q}=
\frac{1}{\sqrt{2}}\sqrt{\sqrt{1+\frac{4}{(\alpha\tau_Q)^2}}-1}.
\end{equation}
Dynamics of the system is governed by the Schr\"odinger equation:
$$i \frac{d}{dt}|\Psi\rangle=H|\Psi\rangle,$$
and it will be assumed that  evolution happens in the interval $[t_i,t_f]$.

As was discussed in \cite{bodzio}, by using AI approximation  one can easily arrive
at the following predictions for the probability of finding the system 
in the excited state at $t_f\gg\hat{t}$  
\begin{itemize}
\item time evolution starts at $t_i\ll-\hat{t}$ from the ground state
\begin{equation}
\label{infty}
P_{AI}=|\langle\uparrow(\hat{t}\,)|\downarrow(-\hat{t}\,)\rangle|^2=
\frac{\hat{\varepsilon}^2}{1+\hat{\varepsilon}^2}.
\end{equation}
In this case the system undergoes in the AI scheme the adiabatic time
evolution from $t_i$ to $-\hat{t}$, then impulse one from $-\hat{t}$
to $\hat{t}$ and finally adiabatic from $\hat{t}$ to $t_f\gg\hat{t}$.
\item time evolution starts at $t_i=0$ from the ground state
\begin{equation}
\label{srodek}
p_{AI}=|\langle\uparrow(\hat{t}\,)|\downarrow(0)\rangle|^2=
\frac{1}{2}\left(1-\frac{1}{\sqrt{1+\hat{\varepsilon}^2}}\right).
\end{equation}
Now  evolution is first impulse from  $t=0$ to $t=\hat{t}$
and then adiabatic from $\hat{t}$ to $t_f\gg\hat{t}$.
\end{itemize}
In the following we  consistently denote excitation probability when the
system evolves from $t_i\ll-\hat{t}$, $t_i=0$ by $P$, $p$ respectively.
Additionally the subscript AI will be attached to predictions based on  
the AI approximation.

In the first case,  the substitution 
of (\ref{hatt}) into (\ref{infty}) leads to 
\begin{equation}
P_{AI}=1- \alpha\tau_Q+\frac{(\alpha\tau_Q)^2}{2}-\frac{(\alpha\tau_Q)^3}{8}+
{\cal O}(\tau_Q^4).
\label{dupa1}
\end{equation}
Now we have to determine $\alpha$. It turns out that it can be
done  by looking at the diabatic excitation probability. The proper 
expression and calculation can be found in Appendix \ref{a_l}. Substituting 
$\eta=1/2$ into (\ref{alfa_1}) and (\ref{infty_1})
one gets that to the lowest nontrivial order $P_{AI}=1-\pi\tau_Q/2$,
which implies that $\alpha=\pi/2$ and  verifies  Eq.
 (\ref{bodzik})
for this case. Note that the calculation leading to Eqs.
(\ref{infty_1}) and (\ref{alfa_1}) is not only much easier then determination 
of exact LZ solution \cite{zener}, but also really elementary. Therefore,
we expect that it can be done  comparably easily for any model of
interest.

\begin{figure}
\includegraphics[width=\columnwidth, clip=true]{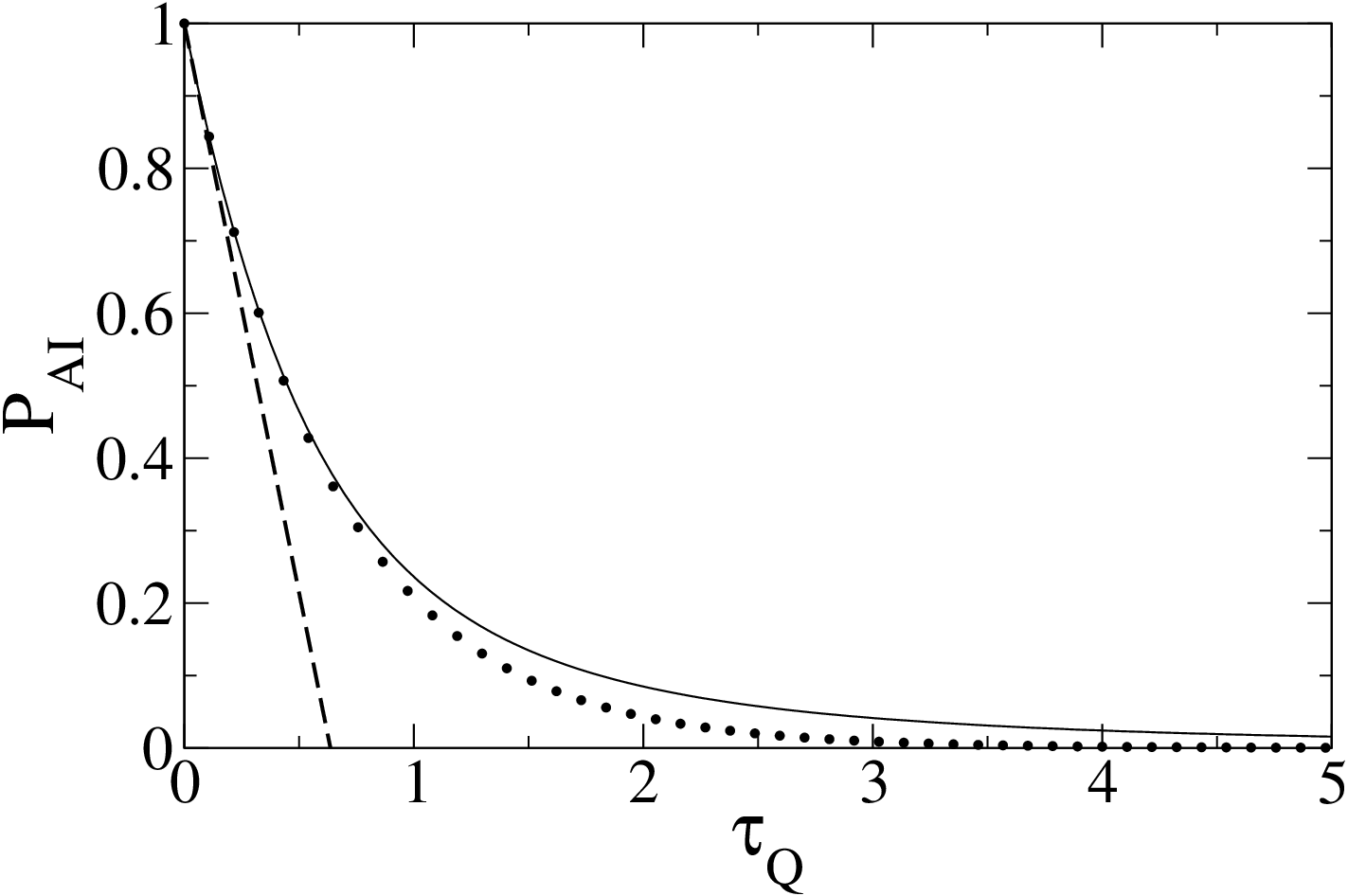}
\caption{Transition probability when the system starts time evolution from a
ground state 
at $t_i\to-\infty$ and evolves to $t_f\to+\infty$.
Dots: exact expression (\ref{dupa987}). Solid line: AI prediction
(\ref{infty_ful}) with $\alpha=\pi/2$ determined from diabatic solution
in Appendix \ref{a_l}. Dashed thick line:  lowest order diabatic result,
$1-\pi\tau_Q/2$, coming from (\ref{infty_1}) and (\ref{alfa_1}) with $\eta=1/2$.
}
\label{dupa123}
\end{figure}

Now we are ready to compare our AI approximation with $\alpha$ determined as
above, to the exact result, i.e.,
\begin{equation}
P=\exp\left(-\frac{\pi\tau_Q}{2}\right).
\label{dupa987}
\end{equation}
First, the
agreement between the exact and AI result is  up to ${\cal O}(\tau_Q^3)$,
i.e., one order above the first nontrivial term. This is the advantage that the 
AI approximation provides over a simple diabatic approximation performed in
Appendix \ref{a_l}.
Second, we see that the AI expansion contains the same powers of $\tau_Q$
as the diabatic (small $\tau_Q$) expansion of the exact result. Third,
Fig. \ref{dupa123} quantifies the discrepancies between exact, AI and diabtic
results. For the AI prediction, we plot in Fig. \ref{dupa123}
instead of a Taylor series (\ref{dupa1})  the full expression evaluated in
\cite{bodzio}
\begin{equation}
P_{AI}= \frac{2}{(\alpha\tau_Q)^2+\alpha\tau_Q\sqrt{(\alpha\tau_Q)^2+4}+2},
\label{infty_ful}
\end{equation}
with $\alpha=\pi/2$. As easily seen the AI approximation significantly
outperforms a diabatic solution. In other words, the combination of 
AI simplification of  dynamics and diabatic prediction for the purpose of
getting the constant $\alpha$ leads to fully satisfactory results considering
simplicity of the whole approach.

It is instructive to consider now separately three situations: (i) dynamics 
in a nonsymmetric avoided level crossing (Sec. \ref{LZa}); (ii) dynamics 
beginning at the anti-crossing center (Sec. \ref{LZb});
and (iii) dynamics starting at $t_i\to-\infty$ but ending at the anti-crossing center
(Sec. \ref{LZc}).
The first case will give us a hint whether  ${\cal O}(\tau_Q^3)$
agreement we have seen above is accidental and has something to do with the symmetry of the
Landau-Zener problem. The second problem 
was preliminarily considered in \cite{bodzio}, but without
comparing the AI prediction to exact analytic one being  interesting on its own. 
Finally, the third problem
is an example where AI approximation correctly suggests  at
a first sight unexpected symmetry  between this problem and the one considered in Sec.
\ref{LZb}.

\subsection{Nonsymmetric Landau-Zener problem}
\label{LZa}

\begin{figure}
\includegraphics[width=\columnwidth, clip=true]{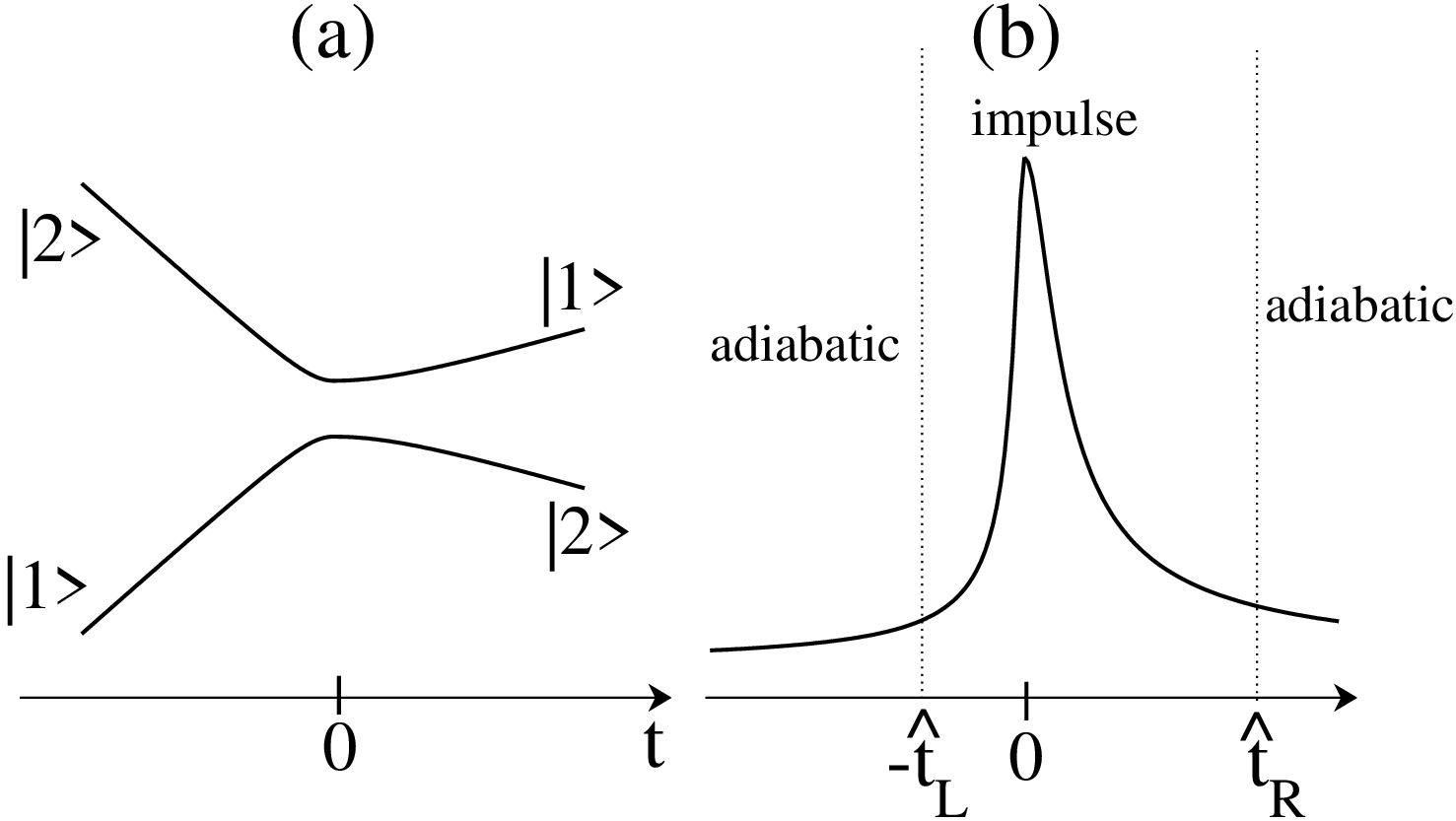}
\caption{The same as in Fig. \ref{gapanty} but for nonsymmetric Landau-Zener 
problem with $\delta>1$ -- see (\ref{Hdelta}).}
\label{asym}
\end{figure}

We assume  that system Hamiltonian  is provided by
the following expression 
\begin{equation}
H=\frac{1}{2}
\left(
\begin{array}{cc}
\frac{1}{\chi}\frac{t}{\tau_Q} & 1 \\
1 & -\frac{1}{\chi}\frac{t}{\tau_Q}
\end{array}
\right) \ \ \ , \ \ \ 
\chi= \left\{ 
\begin{array}{c}
1 \ {\rm for} \ t\le0 \\
\delta \ {\rm for} \ t>0 \\
\end{array}
\right.
\label{Hdelta}
\end{equation}
with $\delta>0$ being the asymmetry parameter -- see Fig. \ref{asym}(a) for
schematic plot of the spectrum.

Once again, evolution starts at $t_i\to-\infty$ from a ground state.
The exact expression for  finding the system
in the excited eigenstate at the end of time evolution ($t_f\to+\infty)$ is
\begin{eqnarray}
\label{exact_nonsym}
P&=&1-
\frac{e^{-\frac{1}{8}\pi(1+\delta)\tau_Q}}{2}
\sinh\left(\frac{1}{4}\pi\tau_Q\delta\right)\nonumber\\&&
\left|\frac{\Gamma(1/2+i\tau_Q\delta/8)}{\Gamma(1/2+i\tau_Q/8)}+
\sqrt{\frac{1}{\delta}}
\frac{\Gamma(1+i\tau_Q\delta/8)}{\Gamma(1+i\tau_Q/8)}\right|^2,
\end{eqnarray}
and its derivation is presented in  Appendix \ref{a_lz}.
Naturally, for $\delta=1$, i.e., in a symmetric LZ problem,
the expression (\ref{exact_nonsym}) reduces to (\ref{dupa987}).

Now we would like to compare (\ref{exact_nonsym}) to predictions coming from
AI approximation. Due to asymmetry of the Hamiltonian the systems enters 
the impulse regime in the time interval $[-\hat{t}_L,\hat{t}_R]$ 
-- see Fig. \ref{asym}(b)
for illustration of these concepts. 
The instants $\hat{t}_L$ and $\hat{t}_R$
are easily found in the same way as in the symmetric case.  
It is a straightforward exercise to verify that according to
AI approximation, the probability of finding the system in the upper state is
\begin{equation}
\label{nonsym}
P_{AI}= |\langle \downarrow(-\hat{t}_L)|\uparrow(\hat{t}_R)\rangle|^2.
\end{equation}

To simplify comparison between exact (\ref{exact_nonsym}) and approximate
(\ref{nonsym}) excitation probabilities we will present their diabatic Taylor expansions:
\begin{eqnarray}
P_{AI}&=& 1- \frac{1}{4}\left( 1+\sqrt{\delta}\right)^2 \alpha\tau_Q +
\nonumber\\&&
\frac{1}{16}(1+\delta)(1+\sqrt{\delta})^2 (\alpha\tau_Q)^2+
 {\cal O}(\tau_Q^3).
 \label{dupa}
\end{eqnarray}
Determination of the constant $\alpha$ is easy and is presented in
Appendix \ref{a_l}. By putting $\eta=1/2$ into (\ref{alfa_1}) and (\ref{2nasym}), one
gets that $\alpha=\pi/2$.  

The AI prediction can be easily compared to the exact result (\ref{exact_nonsym}) 
after series expansion 
$$
P=
1- \frac{\pi}{8}\left( 1+\sqrt{\delta}\right)^2 \tau_Q +
\frac{\pi^2}{64}(1+\delta)(1+\sqrt{\delta})^2 \tau_Q^2
+ {\cal O}(\tau_Q^3).
$$
This comparison shows that once again there is  perfect matching between exact and
AI description up to ${\cal O}(\tau_Q^3)$, i.e., one order better then the simple
diabatic approximation from Appendix \ref{a_l}, despite the fact that we deal
now with nonsymmetric LZ problem.
Moreover, the constant $\alpha$ is the same in symmetric and nonsymmetric cases.
Finally the same powers of $\tau_Q$ show up in both exact and AI results.

The nonsymmetric Landau-Zener model is also interesting in the light of 
a recent paper \cite{ray}, where it is argued that the final state of the
system that passed through a classical
phase transition point, is
determined by details of dynamics after phase transition.
In our system, there are  two different quench rates: $\tau_Q$ 
before the transition and $\tau_Q'=\tau_Q\delta$ after the transition.
Substitution of $\delta=\tau_Q'/\tau_Q$ into 
(\ref{exact_nonsym}) shows that the final state of nonsymmetric Landau-Zener
model expressed in terms of the excitation probability depends on both $\tau_Q$ and
$\tau_Q'$, so that it behaves differently than the system described in
\cite{ray}. 

\subsection{Landau-Zener problem when time evolution starts at the
anti-crossing center}
\label{LZb}
Now we consider the LZ problem characterized by Hamiltonian (\ref{H}) in the
case when time evolution starts from the ground state at the anticrossing
center, i.e., $t_i=0$. It means that 
$|\Psi(0)\rangle\propto\left(|1\rangle-|2\rangle\right)/\sqrt{2}$.
Excitation probability at $t_f\to+\infty$ equals exactly (see Appendix \ref{a_lz})
\begin{eqnarray}
p&=& 1- \frac{2}{\pi\tau_Q} \sinh\left(\frac{\pi\tau_Q}{4}\right)e^{-\pi\tau_Q/8}
   \left|\Gamma\left(1+\frac{i\tau_Q}{8}\right)+\right. \nonumber\\&&\left.
   e^{i\pi/4}\sqrt{\frac{\tau_Q}{8}}
   \Gamma\left(\frac{1}{2}+\frac{i\tau_Q}{8}\right)\right|^2.
\label{srodek_ex}
\end{eqnarray}

From the point of view of AI approximation the evolution is now simplified by
assuming that it is impulse from $t_i=0$ to $\hat{t}$ and then adiabatic from
$\hat{t}$ to $t_f\to+\infty$. This case was discussed in \cite{bodzio}, and it
was shown that substitution of (\ref{hatt}) into (\ref{srodek}) leads to a
simple expression 
\begin{equation}
p_{AI}=
\frac{1}{2}-\frac{1}{2}\sqrt{1-\frac{2}{(\alpha\tau_Q)^2+
\alpha\tau_Q\sqrt{(\alpha\tau_Q)^2+4}+2}},
\label{ai_full}
\end{equation}
expanding it into diabatic series one gets
\begin{equation}
p_{AI}= \frac{1}{2}-\frac{1}{2}\sqrt{\alpha\tau_Q}+\frac{1}{8}(\alpha\tau_Q)^{3/2}+
   {\cal O}(\tau_Q^{5/2}).
\label{srodek_series_ai}
\end{equation}
Determination of a constant $\alpha$ is straightforward 
(see Appendix \ref{a_l}). Putting $\eta=1/2$ into (\ref{infty_2},\ref{alfa_2})
one easily gets $\alpha=\pi/4$ and confirms that the first nontrivial term 
in (\ref{srodek_series_ai}) is indeed $\propto\sqrt{\tau_Q}$.
This value of $\alpha$ ($\approx0.785$) 
is in agreement with the numerical fit done in \cite{bodzio}
where optimal $\alpha$ was found to be equal to $0.77$. Small disagreement
comes from the fact that now we use just the  
$\frac{1}{2}-\frac{1}{2}\sqrt{\alpha\tau_Q}$ part for getting  $\alpha$,
which is equivalent to making the fit to exact results in the limit of
$\tau_Q\ll1$. In \cite{bodzio} the fit of the whole expression
(\ref{ai_full}) in the range of  $0<\tau_Q<6$ was done. 

Having the exact solution (\ref{srodek_ex}) at hand, we can verify 
rigorously accuracy of (\ref{ai_full},\ref{srodek_series_ai}) with the choice
of $\alpha=\pi/4$. Expanding 
(\ref{srodek_ex}) into diabatic  series  one gets
\begin{equation}
p= \frac{1}{2}-\frac{\sqrt{\pi}}{4}\sqrt{\tau_Q}+ 
   \frac{\pi^{3/2}}{64}\left(2-\frac{4\ln2}{\pi}\right)\tau_Q^{3/2}+
   {\cal O}(\tau_Q^{5/2}).
\label{srodek_series}
\end{equation}
As expected the first nontrivial term is in perfect agreement with (\ref{srodek_series_ai})
once $\alpha=\pi/4$. On the other hand, the term
proportional to $\tau_Q^{3/2}$ is about $10\%$ off. Indeed, 
after substitution $\alpha=\pi/4$ into
(\ref{srodek_series_ai}) one gets the third term equal to
$\pi^{3/2}\tau_Q^{3/2}/64$, which has to be compared to 
$\approx1.1\pi^{3/2}\tau_Q^{3/2}/64$ from (\ref{srodek_series}).
It shows that the AI approximation in this case does not predict exactly 
higher nontrivial terms in the diabatic expansion. 

\begin{figure}
\includegraphics[width=\columnwidth, clip=true]{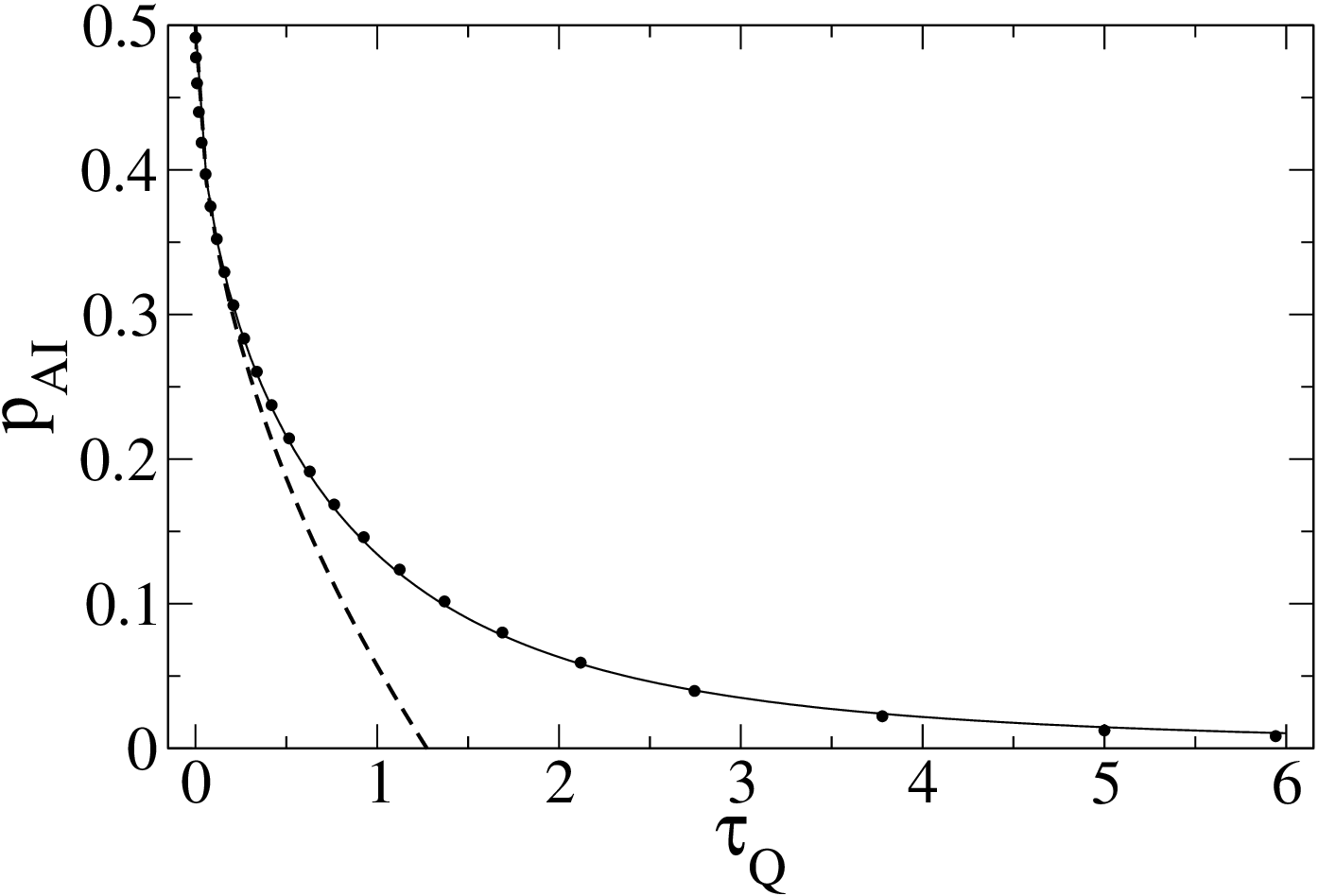}
\caption{Transition probability when the system starts time evolution 
at the anti-crossing center from a ground state and evolves to $t_f\to+\infty$.
Dots: exact expression (\ref{srodek_ex}). Solid line: AI prediction
(\ref{ai_full}) with $\alpha=\pi/4$ determined from the  diabatic solution
in Appendix \ref{a_l}. Dashed thick line: lowest order diabatic prediction
$1/2-\sqrt{\pi\tau_Q}/4$ coming from (\ref{infty_2},\ref{alfa_2}) with $\eta=1/2$.
}
\label{srodek_plot}
\end{figure}

Nonetheless, expression (\ref{ai_full}) works  much better then the above  
comparison suggests. It not only significantly outperforms the 
lowest order diabatic expansion,
$\frac{1}{2}-\frac{1}{4}\sqrt{\pi\tau_Q}$, 
but also beautifully fits to the exact result pretty far from the $\tau_Q\ll1$
regime.
Fig. \ref{srodek_plot} leaves no doubt about these statements.
It suggests that 
 AI predictions should be compared to exact results 
not only term by term as in the diabatic expansion, but also in a full form.

\subsection{One half of Landau-Zener evolution}
\label{LZc}
In this section we consider exactly one half of the LZ problem. Namely,
we take the Hamiltonian (\ref{H}) and evolve the system to the anti-crossing 
center, $t_f=0$, while starting from the ground state at $t_i\to-\infty$.

Let's see what the AI approximation predicts for excitation probability 
of the system at $t_f=0$. According to AI,  evolution is
adiabatic in the interval $[-\infty, -\hat{t}\,]$ and then impulse in time range
$[-\hat{t},0]$.
It implies that
$$|\langle\uparrow(0)|\Psi(0)\rangle|^2\approx
|\langle\uparrow(0)|\Psi(-\hat{t}\,)\rangle|^2\approx|\langle\uparrow(0)|\downarrow(-\hat{t}\,)
\rangle|^2.
$$
A simple calculation then shows that 
\begin{equation}
P_{AI}= |\langle\uparrow(0)|\downarrow(-\hat{t}\,)\rangle|^2=
\frac{1}{2}\left(1-\frac{1}{\sqrt{1+\hat{\varepsilon}^2}}\right).
\label{half_ai}
\end{equation}
A quick look at (\ref{srodek}) shows  that the excitation probability for
one-half of the Landau-Zener problem  is supposed to be, according to the AI scheme, 
equal to the excitation
probability when the system starts from the anti-crossing and evolves toward
$t_f\to+\infty$.

To check that prediction we have solved the one-half LZ model exactly, see
Appendix \ref{a_lz}, and found that indeed both probabilities are exactly 
the same, so the AI approximation provides us here with a  prediction  that is
correct not only qualitatively but also quantitatively.


\section{Dynamics in a general class of quantum two level systems}
\label{2n}
To test predictions coming from the AI approximation on two level systems different 
than the classic
Landau-Zener model, we consider in this section dynamics induced by the Hamiltonian
\begin{equation}
H=\label{H2n}
\frac{1}{2}
\left(
\begin{array}{cc}
{\rm sgn}(t)\left|\frac{t}{\tau_Q}\right|^\frac{\eta}{1-\eta} & 1 \\
1 & -{\rm sgn}(t)\left|\frac{t}{\tau_Q}\right|^\frac{\eta}{1-\eta} 
\end{array}
\right),
\end{equation}
where $\eta\in(0,1)$ is a constant parameter 
(when $\eta=1/2$ the system reduces to the Landau-Zener model).
The eigenstates of Hamiltonian (\ref{H2n}) are expressed by formula (\ref{eig})
with 
$$
\cos(\theta)= \frac{{\rm sgn}(t)\left|\frac{t}{\tau_Q}\right|^\frac{\eta}{1-\eta}}{
              \sqrt{1+\left|\frac{t}{\tau_Q}\right|^\frac{2\eta}{1-\eta}}},
$$
and
$$
\sin(\theta)= \frac{1}{\sqrt{1+\left|\frac{t}{\tau_Q}\right|^\frac{2\eta}{1-\eta}}}.
$$
Extending the notation of Sec. \ref{LZ} to the system (\ref{H2n}) one has to
replace definition (\ref{varepsilon}) by  
$$
\varepsilon= {\rm sgn}(t)\left|\frac{t}{\tau_Q}\right|^\frac{\eta}{1-\eta}
$$
and then expressions (\ref{infty},\ref{srodek}) providing  excitation
probabilities are valid also for the model (\ref{H2n}).

The instant $\hat{t}$ for the Hamiltonian (\ref{H2n}) is determined by 
the following version of Eq. (\ref{bodzik})
\begin{equation}
\label{gap2n}
\frac{1}{\sqrt{1+\left|\frac{\hat{t}}{\tau_Q}\right|^\frac{2\eta}{1-\eta}}}
=\alpha \hat{t}.
\end{equation}
It seems to be impossible to obtain  solution  of (\ref{gap2n}) exactly. Fortunately,
a diabatic perturbative expansion for $\hat{t}$ 
can be  found -- see Appendix \ref{a_r}. Once the
diabatic expansion of $\hat{t}$ is known, we can present different predictions
about dynamics of the model (\ref{H2n}).

\begin{figure}
\includegraphics[width=\columnwidth, clip=true]{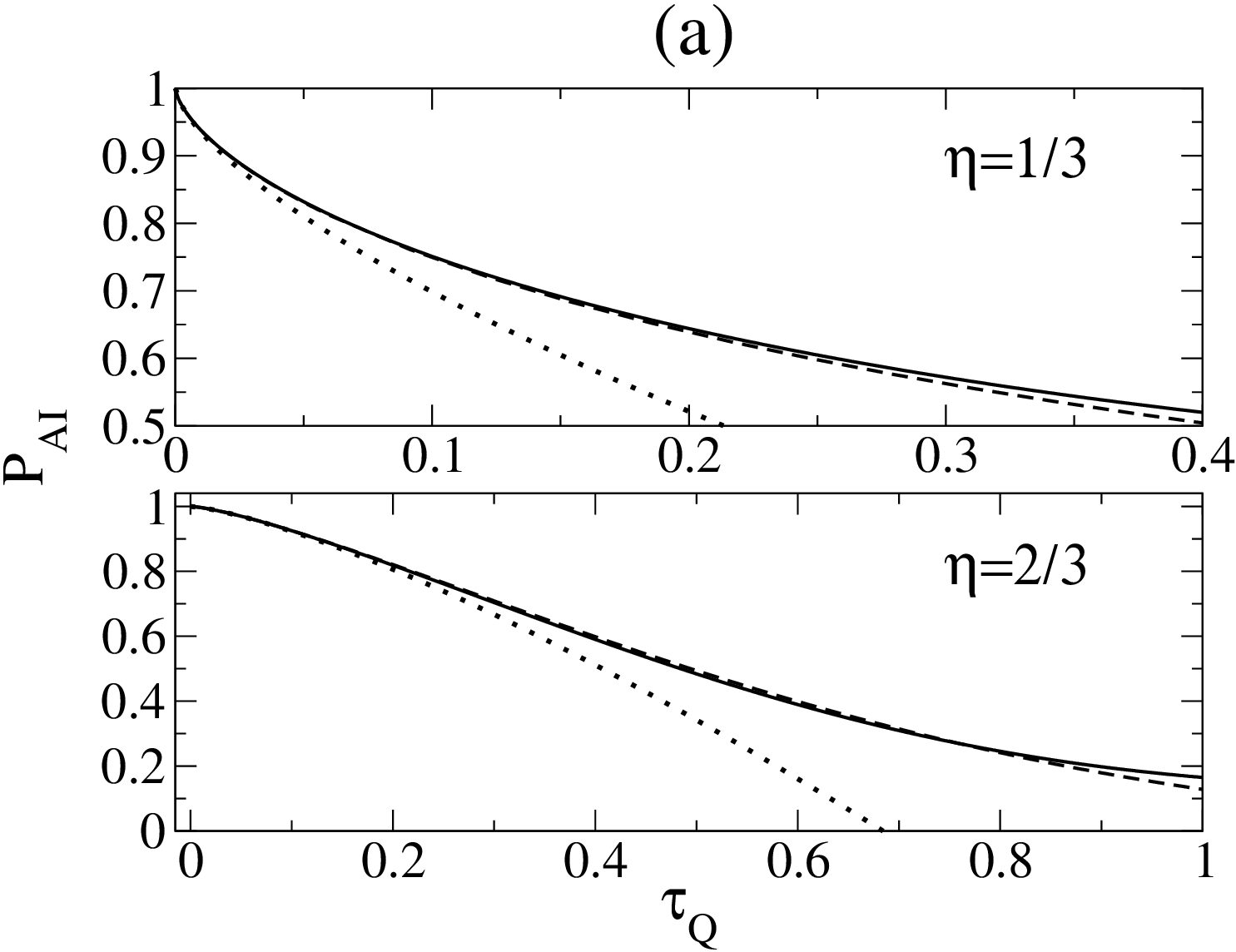}\\
\includegraphics[width=\columnwidth, clip=true]{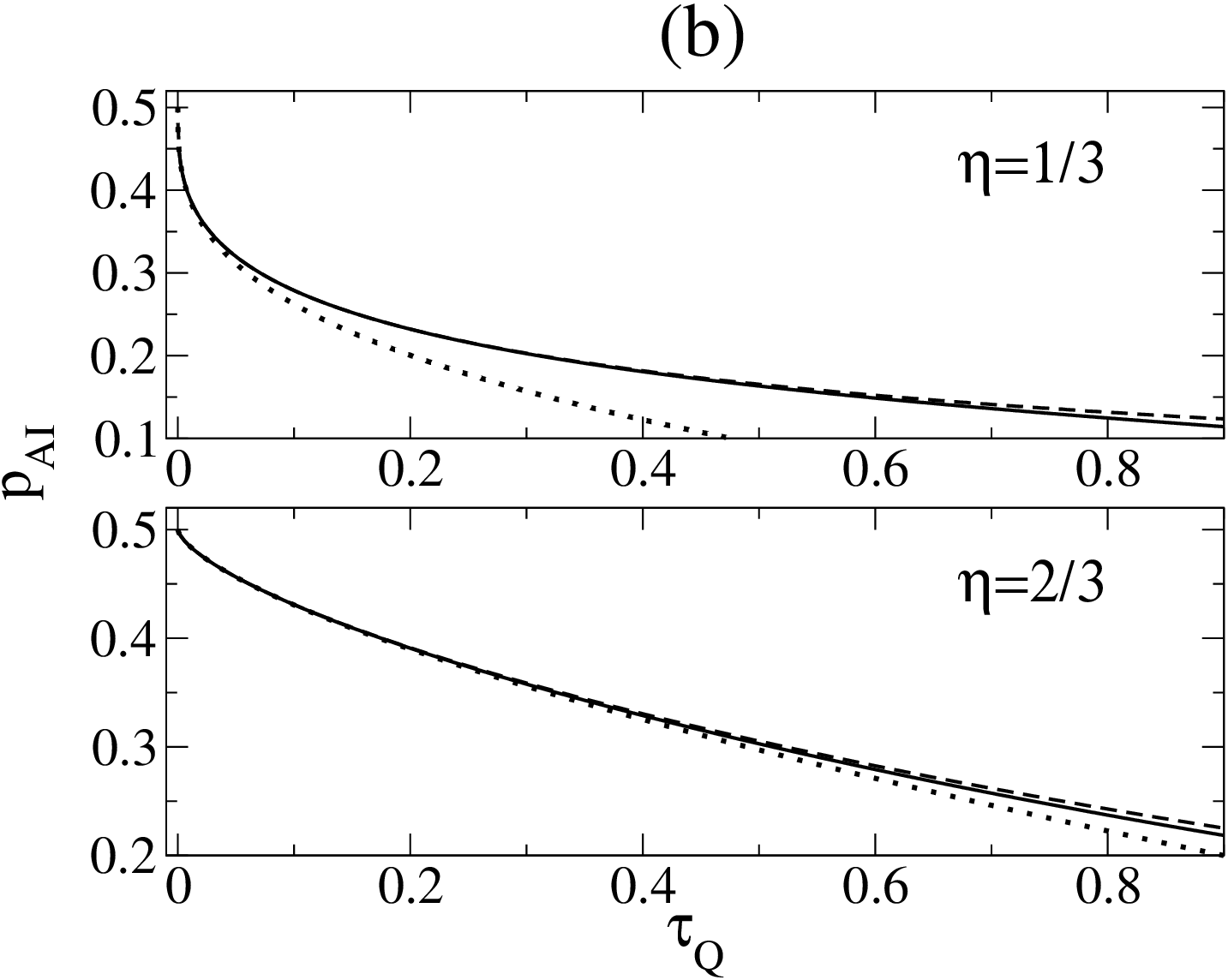}
\caption{Numerics: bold dashed line, AI approximation: solid line, lowest 
order exact diabatic expansion (Appendix \ref{a_l}): dotted line. 
(a): evolution starts from $t_i\to-\infty$.  (b): evolution starts
from $t_i=0$. In both (a) and (b) plots evolution ends in $t_f\to+\infty$.
}
\label{p_excitations}
\end{figure}

First, we consider  transitions starting from a ground state at $t_i\to-\infty$ and
lasting till $t_f\to+\infty$.
Using the results of Appendix \ref{a_r} one gets the following expression for
the excitation probability:
\begin{eqnarray}
P_{AI}&=&\frac{\hat{\varepsilon}^2}{1+\hat{\varepsilon}^2}= 1-(\alpha\tau_Q)^{2\eta}+
(1-\eta)(\alpha\tau_Q)^{4\eta}-\nonumber\\&&
\frac{1}{2}(1-\eta)(2-3\eta)(\alpha\tau_Q)^{6\eta}+\frac{1}{3}(1-\eta)(1-2\eta)
\nonumber\\ & &(3-4\eta)(\alpha\tau_Q)^{8\eta} + 
{\cal O}\left(\tau_Q^{10\eta}\right).
\label{infty2n}
\end{eqnarray}
This result is equivalent to (\ref{dupa1}) when one puts $\eta=1/2$. The constant
$\alpha$ was found in  Appendix \ref{a_l} from the diabatic solution 
and equals (\ref{alfa_1})  
$$\alpha= (1-\eta)\Gamma(1-\eta)^\frac{1}{\eta}.$$
The exponent $2\eta$ in the lowest order  term in (\ref{infty2n})
was positively verified (Appendix \ref{a_l}).
Additionally, we have performed  numerical simulations for $\eta=1/3,2/3, 3/4, 5/6$. 
A fit to numerics in the range of $\tau_Q\ll1$ has  confirmed both the
exponent $2\eta$ and the value of $\alpha$. The comparison between the 
AI prediction for $\eta=1/3,2/3$ and numerics is presented in Fig. \ref{p_excitations}(a).
The lack of exact solution for this problem does not allow us for a more
systematic analytic investigation of the AI approximation in this case.
Nonetheless, a good agreement between AI  and numerics is easily noticed.
Once again, the AI prediction outperforms the lowest order diabatic
expansion.

As before, we would like to consider transitions starting
from the anticrossing center,
i.e., $t_i=0$. In this case the excitation probability equals
\begin{eqnarray}
p_{AI}&=&\frac{1}{2}\left(1-\frac{1}{\sqrt{1+\hat{\varepsilon}^2}}\right)=
      \frac{1}{2}-\frac{1}{2}(\alpha\tau_Q)^\eta+ \frac{1}{4}(1-\eta)\nonumber\\ &&
      (\alpha\tau)^{3\eta}- \frac{1}{16}(1-\eta)(3-5\eta)
      (\alpha\tau_Q)^{5\eta} + \nonumber\\ && {\cal
      O}\left(\tau_Q^{7\eta}\right),
\label{srodek2n}
\end{eqnarray}
As above, the diabatic calculation of Appendix 
\ref{a_l} verifies the exponent of the first nontrivial term in 
(\ref{srodek2n}) and provides us with  the following prediction for $\alpha$
(\ref{alfa_2})
$$
\alpha= (1-\eta)\Gamma(1-\eta)^\frac{1}{\eta}
\cos\left(\frac{\pi}{2}\eta\right)^\frac{1}{\eta}.
$$
After setting $\eta=1/2$  formula (\ref{srodek2n})
becomes the same as (\ref{srodek_series_ai}). Due to the lack of exact results
we have carried out  numerical simulations for $\eta=1/3,2/3,3/4, 5/6$. As expected,
this calculation has  confirmed that the exponent of $\tau_Q$ is indeed
$\eta$ for small $\tau_Q$ and that  $\alpha$ is 
provided by (\ref{alfa_2}). The AI approximation vs. numerics and diabatic 
expansion for $\eta=1/3,2/3$ is presented in Fig. \ref{p_excitations}(b). 
Once again an overall agreement is strikingly good. 

\section{Quantum phase transition in Ising model vs.  adiabatic-impulse
approach}
\label{ising}

In this section we will illustrate how ideas coming from the adiabatic-impulse
approach can be used in studies of dynamics of quantum phase transition. As an
example we  choose the quantum Ising model 
recently considered 
in this context \cite{dorner,jacek,polkovnikov}. 
The quantum Ising model is defined by the following Hamiltonian:
\begin{equation}
\label{His}
H=-\sum_{n=1}^N\left(g\sigma^x_n+\sigma^z_n\sigma^z_{n+1}\right),
\end{equation}
where $N$ is the number of spins. 
The quantum phase transition in this 
model is driven by the change of a dimensionless coupling $g$. The transition point 
between paramagnetic phase ($g>1$) and ferromagnetic one ($0\le g<1$)
is at $g=1$. To study dynamics of (\ref{His}) one assumes
that the system evolves  from time $t'=-\infty$  
to time $t'=0$, and takes
$$
g= -\frac{t'}{\tau_Q'},
$$
where $\tau_Q'$ provides a quench rate. The quantity of interest is
density  of topological defects (kinks) after 
completion of the transition, i.e., at $g=0$. It equals \cite{jacek} 
\begin{equation}
\label{n}
n=\lim_{N\to+\infty}\left\langle\frac{1}{2N}\sum_{n=1}^{N-1}
(1-\sigma^z_n\sigma^z_{n+1})\right\rangle.
\end{equation}
The quantum Ising model obviously possesses $2^N$ different energy 
eigenstates so it seems to be hopeless to expect that the two level
approximation would be sufficient. Therefore, it is a remarkable 
result of Dziarmaga \cite{jacek}, that  dynamics in this system can be 
exactly described by a series of uncoupled Landau-Zener systems.
Due to lack of space, we refer the reader to \cite{jacek},
and present just the main results  and their  AI equivalents.

The density of defects (\ref{n}) in Dziarmaga's notation reads as
$$
n\stackrel{N\gg1}{=}\frac{1}{2\pi}\int_{-\pi}^\pi{\rm d}k\,p_k,
$$
where $p_k$'s are defined as
\begin{equation}
\label{pk}
p_k=|\cos(k/2)u_k(t'=0)-\sin(k/2)v_k(t'=0)|^2,
\end{equation}
with $u_k$, $v_k$ satisfying the following Landau-Zener system
\begin{equation}
\label{lz_ising}
\frac{d}{dt}
\left(
\begin{array}{c}
v_k \\ u_k
\end{array}
\right)
=
\frac{1}{2}
\left(
\begin{array}{cc}
\frac{t}{\tau_Q} & 1 \\
1 & -\frac{t}{\tau_Q}
\end{array}
\right)
\left(
\begin{array}{c}
v_k \\ u_k
\end{array}
\right),
\end{equation} 
$$t=4\left(t'+ \tau_Q'\cos k\right)\sin k,$$
$$\tau_Q= 4\tau_Q'\sin^2k,$$
where  $t'$ changes from $-\infty$
to $0$. The initial conditions for that LZ system are the following 
\begin{equation}
|u_k(t'=-\infty)|=1 \ \ \ , \ \ \ v_k(t'=-\infty)=0.
\label{ukvk}
\end{equation}
Due to the symmetry of the whole problem, it is 
possible to show that $p_k\equiv p_{-k}$, and therefore it is sufficient 
to restrict to $k\ge0$ from now on.

Evolution in (\ref{lz_ising})
lasts from $t_i=-\infty$ to $t_f=2\tau_Q'\sin(2k)$. Since obviously $t_f$
does not go to $+\infty$, it is important to determine where $t_f$ is placed 
with respect to $\pm\hat{t}$. 

We solve the equation $\hat{t}=|t_f|$ (where we substitute $k=k_c$). After 
using (\ref{hatt}) with $\alpha=\pi/2$, and simple algebra one arrives at  
$$
\label{ising_end}
\frac{1}{\sqrt{2}}\sqrt{\sqrt{1+\frac{4}{(\pi\tau_Q/2)^2}}-1}=\left|\frac{\cos
k_c}{\sin k_c}\right|,
$$
which leads to the result 
$$
\sin^2 k_c = 1- \frac{1}{4\pi^2\tau_Q'^{\,2}}. 
$$
Defining $k_c$ to be between $0$ and $\pi/2$ and doing 
some additional easy calculations one gets that 
when $\tau_Q'>\frac{1}{2\pi}$ we have: (i) $t_f\ge\hat{t}$ for $k\in[0,k_c]$;
(ii) $-\hat{t}< t_f < \hat{t}$ for $k\in(k_c,\pi-k_c)$; (iii) 
$t_f\le-\hat{t}$ for $k\in [\pi-k_c,\pi]$.
Moreover, when $\tau_Q'<\frac{1}{2\pi}$ one can show that 
$|t_f|<\hat{t}$ for any $k$'s of interest. It means that the evolution 
ends in the impulse regime for small enough $\tau_Q'$.

Now, as in \cite{dorner,jacek,polkovnikov}, we consider adiabatic time evolutions of the Ising
model. In our scheme
it clearly corresponds to $\tau_Q'\gg\frac{1}{2\pi}$. As the system undergoes
slow evolution it is safe to assume that only long wavelength modes are 
excited, which means that $k\ll\pi/4$ are of interest. This allows 
to approximate  $\sin k\sim k$ and $\cos k\sim1$, which 
implies that (\ref{pk}) turns into $p_k\approx|u_k(t_f)|^2$ and that 
$$
n\approx\frac{1}{\pi}\int_0^{\epsilon}{\rm d}k\,|u_k(t_f)|^2,
$$
with $\epsilon\ll\pi/4$. Since $k\in[0,\epsilon]$ corresponds to the above
mentioned (i) case, $t_f>\hat{t}$ , the AI approximation says that 
$|u_f(t_f)|\approx|u_f(+\infty)|$. Initial conditions (\ref{ukvk}) mean  that 
the system starts time evolution at $t=-\infty$ from {\it excited} state and
we are interested in probability of finding it in the {\it ground} state 
at $t_f$. Elementary algebra based on AI approximation 
shows that this probability equals 
$$
|u_k(t_f)|^2=|\langle \uparrow(-\hat{t}\,)|\downarrow(\hat{t}\,)\rangle|^2=
   |\langle \downarrow(-\hat{t}\,)|\uparrow(\hat{t}\,)\rangle|^2.
$$
Therefore, it is provided by (\ref{infty}) and (\ref{infty_ful}). 
For any fixed $\epsilon$ consistent with lowest order approximation 
of $\sin k$ and $\cos k$ one gets   
\begin{equation}
\label{n_final}
n\cong 0.172\frac{1}{\sqrt{2\tau_Q'}},
\end{equation}
where the prefactor, was found from expansion of the
integral into $1/\sqrt{\tau_Q'}$ adiabatic series with $\tau_Q'\to+\infty$ at fixed $\epsilon$.
In the derivation of this result  $\sin^2 k$ in the  expression for $\tau_Q$ was
approximated by $k^2$. 

First of all, the prediction (\ref{n_final}) 
provides correct scaling
of defect density with $\tau_Q'$. The prefactor, $0.172$, has to be compared to
$\frac{1}{2\pi}\approx0.16$ (exact result from \cite{jacek} and numerical
estimation from \cite{dorner}) or $0.18$ (approximate result from \cite{polkovnikov}).
Our prefactor matches these results  remarkably closely 
concerning simplicity of the whole AI approximation. Slight overestimation 
of the prefactor in comparison to exact result 
comes from the fact that AI transition probability, Eq. (\ref{infty_ful}),
overestimates the exact result, Eq.  (\ref{dupa987}), for large $\tau_Q$'s (Fig. \ref{dupa123}). 
It is interesting, to note that in the Kibble-Zurek scheme used for
description of classical phase transitions an overestimation 
of defect density is usually of the order of  ${\cal O}(1)$
\cite{dorner,antunes,laguna}, while here discrepancy is smaller than $10\%$.

Therefore, the AI approximation provides us with two predictions concerning 
dynamics of the quantum Ising model: (i) it estimates when  evolution ends
in the asymptotic limit; (ii) it correctly
predicts scaling exponent and  number  of defects produced during 
adiabatic transitions. 
The part  (ii)  can be calculated exactly in the quantum Ising model (\ref{His}). 
Nonetheless, in other systems undergoing quantum phase transition 
the two level simplification might be  too difficult for 
exact analytic treatment, e.g., as in the system  governed by (\ref{H2n}). Then 
the AI analysis might be the only analytic approach  working beyond
the lowest order diabatic expansion.
Besides that the AI approach provides us with a quite  intuitive 
description of system dynamics, which is of interest in its own. Especially, when
one looks at connections between classical and quantum phase transitions.

\section{Summary}

We have shown that the adiabatic-impulse approximation,
based on the ideas underlying Kibble-Zurek mechanism \cite{kibble,zurek},
provides good  quantitative predictions concerning 
diabatic dynamics of two level Landau-Zener like systems.
After supplementing the splitting of the evolution 
into adiabatic and impulse regimes by the exact lowest order
diabatic calculation, the whole approach is complete and 
can  be in principle applied to different systems 
possessing anti-crossings, e.g., those lacking exact analytic
solution as the model (\ref{H2n}). We expect that the AI approach 
will provide the link between dynamics of classical and quantum 
phase transition thanks to usage of the same terminology and 
 similar assumptions. 

We also expect that different variations of the classic Landau-Zener system
discussed in this paper can be experimentally realized  
in the setting where the sweep rate can be manipulated.  
Indeed,  the model (\ref{H2n}) 
arises once a proper nonlinear change of external system parameter is 
performed. The experimental access to studies of nonsymmetric Landau-Zener model can be
obtained  by  change of the sweep rate after passing an anti-crossing center.
The case of evolution starting from or ending at the anticrossing center 
can also be subjected to experimental investigations, e.g., in a beautiful
system consisting of the smallest available quantum magnets (cold Fe$_8$
clusters) \cite{magnet}.

We are grateful to Uwe Dorner for collaboration on  nonsymmetric
Landau-Zener problem, and Jacek Dziarmaga for comments and useful 
suggestions on the manuscript.
Work supported by the U.S. Department of Energy,
National Security Agency, and the ESF COSLAB program.


\appendix
\section{Exact diabatic expressions for  transition probabilities}
\label{a_l}
We would like to provide lowest order exact expressions for transition
probabilities in a class of two level systems described by the 
Hamiltonian (\ref{H2n}). 

First, we express the wave function as 
\begin{eqnarray}
|\Psi(t)\rangle&=&
C_1(t)
\exp\left(\frac{-i(1-\eta)|t|^\frac{1}{1-\eta}}{2\tau_Q^\frac{\eta}{1-\eta}}\right)
|1\rangle+\nonumber\\&&
C_2(t)
\exp\left(\frac{i(1-\eta)|t|^\frac{1}{1-\eta}}{2\tau_Q^\frac{\eta}{1-\eta}}\right)
|2\rangle,
\label{psi2n}
\end{eqnarray}
were exponentials are $\mp i\int {\rm dt}\, {\rm sgn}(t)|t/\tau_Q|^\frac{\eta}{1-\eta}/2$.
Within this representation dynamics governed by the Hamiltonian (\ref{H2n})
reduces to 
\begin{eqnarray}
\label{c1nc2n0}
i \dot{C_1}(t)&=& \frac{C_2(t)}{2} 
\exp\left(\frac{i(1-\eta)|t|^\frac{1}{1-\eta}}{\tau_Q^\frac{\eta}{1-\eta}}\right)\\
\label{c1nc2n}
i \dot{C_2}(t)&=& \frac{C_1(t)}{2}
\exp\left(\frac{-i(1-\eta)|t|^\frac{1}{1-\eta}}{\tau_Q^\frac{\eta}{1-\eta}}\right)
\end{eqnarray}

{\bf Time evolution starting from the ground state at $t_i\to-\infty$:}
we would like to integrate (\ref{c1nc2n}) from $-\infty$ to $+\infty$. 
The initial condition is such that $C_1(-\infty)=1$ and $C_2(-\infty)=0$.
The simplification comes when one assumes a very fast transition, i.e.,
$\tau_Q\to0$. Then it is clear that $C_1(t)=1+{\cal O}(\tau_Q^\beta)$ with 
some $\beta>0$. Putting such $C_1(t)$ into (\ref{c1nc2n}) one gets
\begin{eqnarray}
C_2(+\infty)&=& \frac{1}{2i}\tau_Q^\eta\int_{-\infty}^{+\infty} {\rm d}x\,
\exp\left(-i(1-\eta)|x|^\frac{1}{1-\eta}\right)+\nonumber\\&&
({\rm higher \ order \ terms \ in  \ \tau_Q}),
\end{eqnarray}
which after some algebra results in 
\begin{eqnarray}
P=|C_1(+\infty)|^2&=& 1-|C_2(+\infty)|^2= 1-
(\alpha\tau_Q)^{2\eta}\nonumber+\\&& 
({\rm higher \ order \ terms \ in  \ \tau_Q}), 
\label{infty_1}
\end{eqnarray}
where
\begin{eqnarray}
\alpha= (1-\eta)\Gamma\left(1-\eta\right)^\frac{1}{\eta}.
\label{alfa_1}
\end{eqnarray}

{\bf Time evolution starting from the ground state at $t_i=0$ (anti-crossing
center):} since initial wave function is $(|1\rangle-|2\rangle)/\sqrt{2}$ 
we have $C_1(0)=1/\sqrt{2}$ and $C_2(0)=-1/\sqrt{2}$ and we evolve the system
till $t_f\to+\infty$ with the Hamiltonian (\ref{H2n}). For fast transitions one has
that $C_1(t)=1/\sqrt{2}+{\cal O}(\tau_Q^\beta)$, where $\beta>0$ is some
constant. Integrating (\ref{c1nc2n}) from $0$ to $+\infty$ one gets:
\begin{eqnarray}
C_2(+\infty)+\frac{1}{\sqrt{2}}&=& \frac{\tau_Q^\eta}{2\sqrt{2}i}
\int_{0}^{+\infty} {\rm d}x
\exp\left(-i(1-\eta)x^\frac{1}{1-\eta}\right)\nonumber\\&&
+({\rm higher \ order \ terms \ in  \ \tau_Q}),
\end{eqnarray}
which can be easily shown to lead to 
\begin{eqnarray}
p&=&|C_1(+\infty)|^2= 1-|C_2(+\infty)|^2= \frac{1}{2}-\frac{1}{2}
(\alpha\tau_Q)^\eta\nonumber+\\&& 
({\rm higher \ order \ terms \ in  \ \tau_Q}), 
\label{infty_2}
\end{eqnarray}
where
\begin{eqnarray}
\alpha= (1-\eta)\Gamma(1-\eta)^\frac{1}{\eta}
\cos\left(\frac{\pi}{2}\eta\right)^\frac{1}{\eta}.
\label{alfa_2}
\end{eqnarray}

{\bf Time evolution from $t_i\to-\infty$ to $t_f\to+\infty$ in the nonsymmetric LZ model:}
we assume that  the Hamiltonian is given by (\ref{H2n}) for $t\le0$, while for
$t>0$ it is given by (\ref{H2n}) with $\tau_Q$ exchanged by $\tau_Q\delta$
where $\delta>0$ is the asymmetry constant -- see (\ref{Hdelta}) for
$\eta=1/2$ case. Integration of (\ref{c1nc2n})
separately in the intervals $[-\infty,0]$ and $[0,+\infty]$ 
leads to the following prediction
\begin{eqnarray}
\label{2nasym}
P&=&|C_1(+\infty)|^2= 1-|C_2(+\infty)|^2= 1-\frac{1}{4}
\nonumber\\&&\left(1+\delta^\eta\right)^2
(\alpha\tau_Q)^{2\eta}+\nonumber\\&& 
({\rm higher \ order \ terms \ in  \ \tau_Q}), 
\end{eqnarray}
with $\alpha$ given by (\ref{alfa_1}).


\section{Exact solutions of the Landau-Zener system}
\label{a_lz}
In this Appendix we present derivation of exact solutions of the Landau-Zener problem
that are used in the main part of the paper. The general exact solution for Landau-Zener  
model evolving 
according to Hamiltonian (\ref{H}) was first discussed  in
\cite{zener} and then elaborated in a number of papers, e.g.,
\cite{vitanov}. 
Here we will use that general solution 
for getting predictions about different time evolutions.
The problem is simplified by writing the wave function as
(\ref{psi2n}) with $\eta=1/2$. Then  one arrives at the equations 
(\ref{c1nc2n0}) and (\ref{c1nc2n}) once again with $\eta=1/2$. Combining the latter ones
with the substitution 
$$
U_2(t)=C_2(t)e^{\frac{it^2}{4\tau_Q}}
$$
one gets 
$$
\ddot{U}_2(z)+\left(k-\frac{z^2}{4}+\frac{1}{2}\right)U_2(z)=0,
$$
where 
\begin{equation}
\label{k}
k=\frac{i\tau_Q}{4} \ \ , \ \ z=\frac{t}{\sqrt{\tau_Q}}e^{-i\pi/4}.
\end{equation}
The general solution of this equation is expressed in
terms of linearly independent Weber functions $D_{-k-1}(\pm iz)$ \cite{zener,whittaker}. 
By combining this observation with $\eta=1/2$ version of  (\ref{c1nc2n})  one gets
\begin{eqnarray}
|\Psi(t)\rangle&=& 2i\left[\partial_t-\frac{it}{2\tau_Q}\right]
[aD_{-k-1}(iz)+bD_{-k-1}(-iz)]|1\rangle
\nonumber\\&&+
[aD_{-k-1}(iz)+bD_{-k-1}(-iz)]|2\rangle,
\label{general}
\end{eqnarray}
where $z$ is defined in (\ref{k}). Though there is a simple one-to-one correspondence
between $z$ and $t$, we will use both $z$ and $t$ in different expressions to shorten 
notation. 

To determine constants $a$ and $b$ in different cases one has  to know
the following properties of Weber functions \cite{whittaker}.
First, $\forall \ |arg(s)|<3\pi/4$ one has
\begin{eqnarray}
D_m(s)&=& e^{-s^2/4}s^m \left[1+{\cal O}(s^{-2})\right].
\label{as1}
\end{eqnarray}
Second, $\forall -5\pi/4<arg(s)<-\pi/4$ 
\begin{equation}
D_m(s)= e^{-im\pi}D_m(-s)+\frac{\sqrt{2\pi}}{\Gamma(-m)}e^{-i(m+1)\pi/2}D_{-m-1}(is).
\label{as2}
\end{equation}
Third, as $s\to0$ one has
\begin{equation}
D_m(s)=2^{m/2}\frac{\sqrt{\pi}}{\Gamma(1/2-m/2)}+{\cal O}(s).
\label{as0}
\end{equation}
Finally,  
\begin{equation}
\frac{d}{ds}D_m(s)=mD_{m-1}(s)-\frac{1}{2}sD_m(s).
\label{diff}
\end{equation}

{\bf Exact solution of nonsymmetric LZ problem:}
the evolution starts at $t_i\to-\infty$, i.e., up to a  phase factor
$|\Psi(-\infty)\rangle\sim|1\rangle$. The Hamiltonian is given by
(\ref{Hdelta}). Using (\ref{as1},\ref{as2},\ref{diff}) one  finds that  $a=0$ and 
$b=\sqrt{\tau_Q}\exp(-\pi\tau_Q/16)/2$. Substituting them into (\ref{general}) 
results in 
\begin{eqnarray}
|\Psi(t\le0)\rangle &=& e^{-\pi\tau_Q/16}e^{i3\pi/4}\nonumber\\&&
[(k+1)D_{-k-2}(-iz)-izD_{-k-1}(-iz)]|1\rangle\nonumber\\&&
+\frac{\sqrt{\tau_Q}}{2}e^{-\pi\tau_Q/16} D_{-k-1}(-iz)]|2\rangle.
\end{eqnarray}
Using (\ref{as0}) one finds that this solution at $t=0$ becomes 
\begin{eqnarray}
|\Psi(0)\rangle&=& e^{-\pi\tau_Q/16} e^{i3\pi/4} 
\frac{\sqrt{\pi}2^{-k/2}}{\Gamma(1/2+k/2)}|1\rangle+
\nonumber\\&&  
\frac{\sqrt{\tau_Q}}{2}
e^{-\pi\tau_Q/16} 
\sqrt{\frac{\pi}{2}}\frac{2^{-k/2}}{\Gamma(1+k/2)}  |2\rangle.
\label{t0}
\end{eqnarray}
For $t>0$ the Hamiltonian changes its form, see (\ref{Hdelta}), and one has to match 
(\ref{t0}) with (\ref{general}) having $\tau_Q$ replaced by
$\tau_Q\delta$. Substitution of 
\begin{eqnarray}
a&=&\frac{e^{-\pi\tau_Q/16}}{4}2^{-k(1-\delta)/2} \sqrt{\tau_Q\delta}
\left[\frac{\Gamma(1+k\delta/2)}{\Gamma(1+k/2)}\sqrt{\frac{1}{\delta}} \right.
\nonumber\\&&\left.
-\frac{\Gamma(1/2+k\delta/2)}{\Gamma(1/2+k/2)}\right],\nonumber
\end{eqnarray}
\begin{eqnarray}
b&=&\frac{e^{-\pi\tau_Q/16}}{4}2^{-k(1-\delta)/2} \sqrt{\tau_Q\delta}
\left[\frac{\Gamma(1+k\delta/2)}{\Gamma(1+k/2)}\sqrt{\frac{1}{\delta}} \right.
\nonumber\\&&\left.
+\frac{\Gamma(1/2+k\delta/2)}{\Gamma(1/2+k/2)}\right],\nonumber
\end{eqnarray}
into (\ref{general}) gives the wave function $|\Psi(t\ge0)\rangle$.
Taking one minus squared modulus of the amplitude of finding the system in the state
$|2\rangle$ one obtains the excitation probability of the system at
$t_f\to+\infty$ in the form (\ref{exact_nonsym}).

{\bf Exact solution of LZ problem when evolution starts from a ground state at
anticrossing center:}
the Hamiltonian of the system is given by (\ref{H}) and we look for a solution
that starts at $t=0$ from the (ground) state:
$|\Psi(0)\rangle=(|1\rangle-|2\rangle)/\sqrt{2}$. The constants $a$
and $b$ from (\ref{general}) turn out to be equal to 
$$
a= \frac{2^{k/2}\Gamma(1/2+k/2)e^{i\pi/4}\sqrt{\tau_Q}}{4\sqrt{2\pi}}
-2^{k/2}\frac{\Gamma(1+k/2)}{2\sqrt{\pi}},
$$
$$
b= -\frac{2^{k/2}\Gamma(1/2+k/2)e^{i\pi/4}\sqrt{\tau_Q}}{4\sqrt{2\pi}}
-2^{k/2}\frac{\Gamma(1+k/2)}{2\sqrt{\pi}}.
$$
Having this at hand,  
it is straightforward to show that one minus squared modulus of the 
amplitude of finding the system in the state $|2\rangle$ at $t_f\to+\infty$, 
i.e., excitation probability, equals (\ref{srodek_ex}).

{\bf One-half of the Landau-Zener problem:}
the excitation probability of the system 
that started time evolution at $t_i=-\infty$ from ground state,
and stopped evolution at $t_f=0$,
turns out to be equal to  (\ref{srodek_ex}).
The simplest way to prove it comes from an observation that this
probability equals one minus 
the squared overlap between the state $(|1\rangle-|2\rangle)/\sqrt{2}$
and (\ref{t0}). Then, straightforward calculation leads immediately to 
expression (\ref{srodek_ex}).

\section{Solution of Eq. (\ref{gap2n})}
\label{a_r}

\begin{figure}[t]
\includegraphics[width=\columnwidth, clip=true]{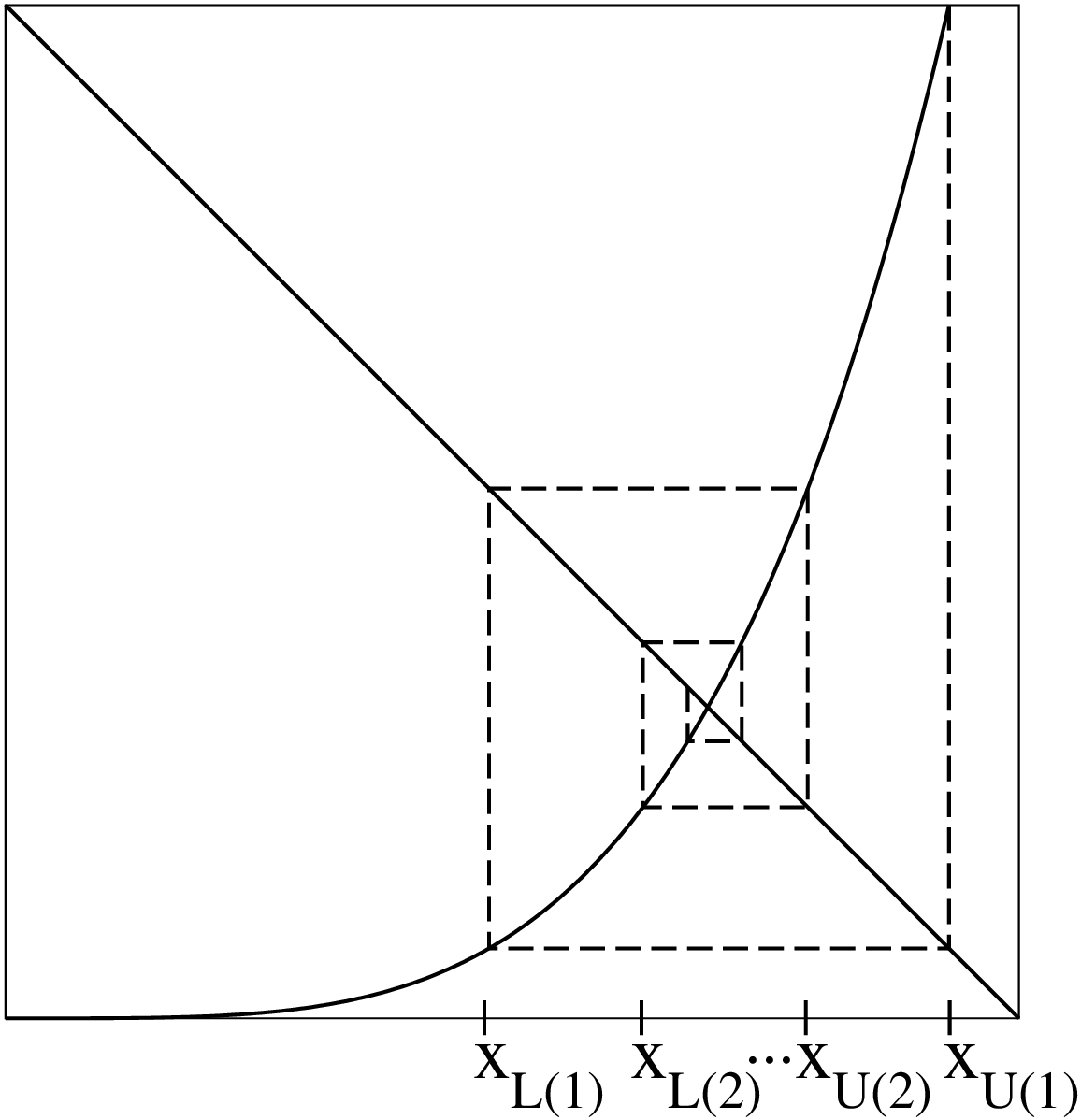}
\caption{Schematic plot of the recurrence method for getting solution of 
(\ref{gap2n_new}). Solid lines $x^\frac{1}{1-\eta}$ and $\beta-x$, dashed line is a
construction of a recurrence solution  (\ref{rec1}) and (\ref{rec2}). 
The plots are in a $\beta\times\beta$ box.}
\label{rec}
\end{figure}

Equation (\ref{gap2n}) can be solved in the following recursive way.
First, one introduces new variables:
\begin{equation}
\label{variables}
x=\left(\frac{\hat{t}}{\tau_Q}\right)^2 \ \ , \ \ \beta=\frac{1}{\alpha^2\tau_Q^2}.
\end{equation}
In these variables Equation (\ref{gap2n}) becomes:
\begin{equation}
\label{gap2n_new}
x^\frac{1}{1-\eta}= \beta-x,
\end{equation}
and we assume that $\beta>1$ (diabatic evolutions).

A quick look at  Fig. \ref{rec}, makes clear that the solution can
be obtained by considering a series of inequalities, numbered by index $i$,
in the form
\begin{equation}
x_{L(i)} < x < x_{U(i)},
\label{rec1}
\end{equation}
where  both lower ($x_{L(i)}$) and upper ($x_{U(i)}$) bounds 
satisfy the the same recurrence equation
\begin{equation}
x_{L,U(i)}= \left[\beta -
\left[\beta-x_{L,U(i-1)}\right]^{1-\eta}\right]^{1-\eta},
\label{rec2}
\end{equation}
with the initial conditions $x_{L(0)}=0$, $x_{U(0)}=\beta$. Considering a few
iterations one can show that the solution of (\ref{gap2n_new}) can be
conveniently written as 
\begin{eqnarray}
\label{xdelta}
x^\frac{1}{1-\eta} &=& \frac{1}{(\alpha\tau_Q)^2}\left[1-
(\alpha\tau)^{2\eta}+
(1-\eta)(\alpha\tau_Q)^{4\eta}+\right.\nonumber\\   
& &  \frac{1}{2}(3\eta-2)(1-\eta)(\alpha\tau_Q)^{6\eta}+ 
\frac{1}{3}(1-\eta)(1-2\eta)\nonumber\\ & &(3-4\eta) \left.
(\alpha\tau_Q)^{8\eta}+{\cal O}\left(\tau_Q^{10\eta}\right)\right].
\end{eqnarray}
Using (\ref{xdelta}), different quantities of interest can be determined,
e.g., 
\begin{eqnarray}
\label{epsilon2}
\hat{\varepsilon}^2&=&\frac{1}{(\alpha\tau_Q)^{2\eta}}\left[1-\eta
(\alpha\tau_Q)^{2\eta}+
\frac{1}{2}\eta(1-\eta)(\alpha\tau_Q)^{4\eta}- \right.\nonumber\\
&&\frac{1}{3}\eta(1-\eta)(1-2\eta)(\alpha\tau_Q)^{6\eta}+
\frac{1}{8}\eta(1-\eta)\nonumber\\ & & \left. (1-3\eta)(2-3\eta)(\alpha\tau_Q)^{8\eta}+
{\cal O}\left(\tau_Q^{10\eta}\right)\right],\\
\label{that2n}
\hat{t}&=& \tau_Q^\eta\alpha^{\eta-1}+ {\cal
O}\left(\tau_Q^{3\eta}\right).
\end{eqnarray}
Substitution of $\eta=1/2$ into (\ref{epsilon2}) and (\ref{that2n}) reproduces 
diabatic series expansions of $\hat{\varepsilon}$, $\hat{t}$ from the standard
Landau-Zener theory (\ref{hatt}).

\end{document}